\documentclass[twocolumn, preprintnumbers,amsmath,amssymb, prb,floatfix]{revtex4}
\usepackage{graphicx}
\usepackage{dcolumn}
\usepackage{bm}
\usepackage{float}
\usepackage[T1]{fontenc}
\usepackage{color}
\usepackage{epstopdf}

\usepackage{graphics}

\begin{document}

\preprint{APS/123-QED}

\title{Multiple charge density wave transitions in Lu$_{2}$Ir$_{3}$Si$_{5}$
single crystal }

\author{N. S. Sangeetha$^{1}$, A. Thamizhavel$^{2}$, C. V. Tomy$^{3}$,
Saurabh Basu$^{1}$, A. M. Awasthi$^{4},$ S.~Ramakrishnan$^{2}$
and D. Pal$^{1}$}

\email{dpal@iitg.ernet.in}

\affiliation{$^{1}$Department of Physics, Indian Institute of Technology Guwahati,
Guwahati, Assam-781039, India }

\affiliation{$^{2}$Department of Condensed Matter Physics and Materials Science,
Tata Institute of Fundamental Research, Homi Bhabha Road, Colaba,
Mumbai-400 005, India}

\affiliation{$^{3}$Department of Physics, Indian Institute of Technology Bombay,
Mumbai - 400076, India}

\affiliation{$^{4}$IUC-DAEF, Indore-452017, India}

\date{\today}
\begin{abstract}
The physical properties of the single-crystalline samples of Lu$_{2}$Ir$_{3}$Si$_{5}$
have been investigated by magnetic susceptibility, resistivity and
heat capacity studies. We observed multiple
charge density wave (CDW) transitions in all the measurements. A strong
thermal hysteresis at these transitions suggests a possible first order CDW ordering.
In addition, the first order nature is ascertained by a very narrow and
a huge cusp (62~J/mol~K) in the zero field specific heat data which
also suggests a strong interchain coupling. By applying a field of
9T in the specific heat measurement, one of the CDW transitions is
suppressed.
\end{abstract}
\maketitle

\section{Introduction}

Charge-density-wave (CDW) transitions expected to occur  in low dimensional solids
where it is possible to achieve nesting of Fermi surfaces that leads
to the appearance of a periodic lattice distortion with an accompanying
energy gap. 
The instability to achieve the instability in low dimensional solids
was first demonstrated by Pierels \cite{Peierls_book,Gruner}.
This is very well documented from the early works
of many research groups performed on a wide range of quasi-low-dimensional
systems includes transition-metal dichalcogenides and trichalcogenides
\cite{KMoO3,Nb_Tb_Se3_1,Bechgaard_salt,TTF-TCNQ2}. Even though in
three-dimensional (3D) compounds it is not possible to get a perfect
nesting but CDW ordering has also been established in 3D materials
such as  R$_{2}$Ir$_{3}$Si$_{5}$\cite{R2Ir3Si5_Yogesh_18}
and R$_{5}$Ir$_{4}$Si$_{10}$ \cite{R5Ir4Si10_14} (\textit{R}~=~rare-earth
elements). So it appears that even in the presence of imperfect nesting there remains
a possibility for the appearance of a CDW. In order to understand the nature of
such novel CDW in such systems, a further search for new classes of materials
are needed. 

Recently reported R$_{5}$Ir$_{4}$Si$_{10}$ (R=Dy-Lu)
compounds exhibit strong coupling CDW at high temperatures accompanying
with superconductivity or magnetic ordering at low temperature \cite{R5Ir4Si10_15}.
Interestingly, the multiple CDW anomalies were observed in Dy, Ho,
Er and Tm \cite{multipleCDW_5-4-10}. The compound Lu$_{5}$Ir$_{4}$Si$_{10}$
shows the coexistence of superconductivity and strongly coupled CDW
transition \cite{SCrystal_Lu5Ir4Si10}. From the literature, it is
found that the compound R$_{5}$Ir$_{4}$Si$_{10}$ presents a complex
3D structure with several substructures such as one-dimensional R
chains and 3D cages in which a variety of many-body effects (superconductivity
and magnetism) originate. Similar complex behavior has recently reported
in R$_{2}$Ir$_{3}$Si$_{5}$ system, which display various unusual
ground states like, superconductivity, CDW, Kondo behavior, coexistence
of CDW and superconductivity or magnetism, etc . 

In R$_{2}$Ir$_{3}$Si$_{5}$ series, the compound
Lu$_{2}$Ir$_{3}$Si$_{5}$ is of special interest, as it exhibits
superconductivity at 3.5~K and shows strongly coupled first order
CDW transition between 150 and 200~K.\cite{lu2Ir3Si5_yogesh}.
Kuo \textit{et al}. recently reported the possibility for the CDW
transition accompanying with a similar structural phase transition
followed by the transport measurements \cite{Lu2r3Si5_3}. Besides,
in our earlier work, we have reported a non-monotonicity
of the transition temperatures, \textit{T}{$_{CDW}$
and \textit{T}{$_{SC}$ as a
function of Ge substitution \cite{Sangu}. This study revealed the complex nature
of the CDW ordering such that the system may undergoes coexistence/mixture of multiple phase
CDW ordering with the suppression of CDW upto certain Ge concentration afterwards a 
sudden enhancement of it.
Definite understanding of this feature needs further experimentation
on  single crystal of Lu$_{2}$Ir$_{3}$Si$_{5}$.  In this paper we provide
an observation of multiple CDW transitions in Lu$_{2}$Ir$_{3}$Si$_{5}$
single crystals using thermodynamic, transport and magnetic
measurements.

\section{Experiment}

Two single-crystalline samples of Lu$_{2}$Ir$_{3}$Si$_{5}$ were
grown in a tetra-arc furnace using a modified Czochralski technique.
The purity of the elements, melted in a stoichiometric ratio, were
Lu: 99.999\%; Ir: 99.99\%; Si: 99.999\%. The samples were characterized
using scanning electron microscopy equipped with energy-dispersive
X-ray analysis (EDAX) to prove its 2-3-5 stoichiometry. The room temperature
powder x-ray diffraction with Cu K$\alpha$ radiation was taken on
the samples by using PANalytical commercial X-ray diffractometer.
A commercial superconducting quantum interference device (SQUID) magnetometer
(MPMS5, Quantum Design, USA) was used to measure dc magnetic susceptibility
as a function of temperature between 100 and 300~K. The electrical
resistivity between 1.8 and 300~K was measured by using a home built
electrical resistivity set up with the standard dc four probe technique.
The specific heat data were taken on the samples for the temperature
range 120 to 300~K in zero field, by using a DSC set up, and in 9~T
field by using a commercial physical property measurement system (PPMS,
Quantum Design, USA). The reproducibility of the
results was checked by repeating the measurements on the same samples.

\section{Results }

\subsection{X-ray diffraction studies}

The powder X-ray diffraction pattern of Lu$_{2}$Ir$_{3}$Si$_{5}$
at 300 K clearly reveals the absence of any impurity phase and also confirms
that the samples have a U$_{2}$Co$_{3}$Si$_{5}$ - orthorhombic
type structure with the space group \textit{Ibam}. The Fullprof Reitveld
fit \cite{Fullprof} to the powder x-ray data of Lu$_{2}$Ir$_{3}$Si$_{5}$
is shown in Fig.~\ref{Fig1_FP}. The extracted lattice parameters
from this fit are $\mathit{a}=\mathrm{9.923}\pm{0.0005}\text{Å}$, $\mathit{b}=\mathrm{11.311}\pm{0.0005}\textrm{Å}$
and $\mathit{c}=\mathrm{5.732\,}\pm{0.0005}\text{Å}$ which are in close agreement
with the previously reported polycrystalline data \cite{lu2Ir3Si5_yogesh}.
The single crystalline nature of the samples was verified by Laue
diffraction technique. 
\begin{figure}[!]
\centering{} \includegraphics[width=8cm,height=8cm,keepaspectratio]{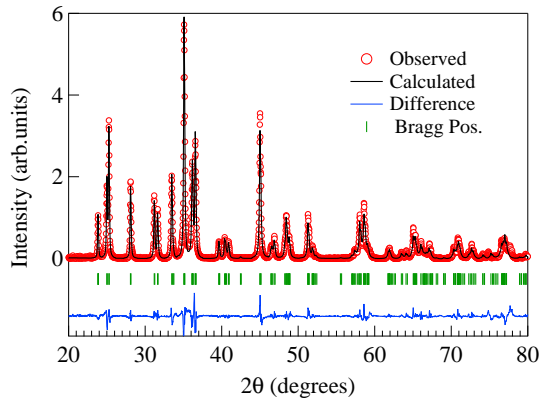}
\caption{\label{Fig1_FP}(color line) Powder X-ray diffraction data of the
Lu$_{2}$Ir$_{3}$Si$_{5}$ at 300 K. The solid line is the simulated data
using the FulProf (Reitveld Program).}
\end{figure}
The single crystals were oriented along three principle
crystallographic directions, which are mutually perpendicular to each
other, by using back-reflection Laue diffraction method. The observed
Laue pattern of each direction was analyzed, with the simulated pattern,
by using Orient Express software. The observed Laue pattern of an
oriented crystal, along {[}100{]} direction, is shown in Fig.~\ref{Fig2_LP}.
For transport and magnetization measurements, small bars (required
size) were cut by spark erosion from the oriented crystals. 

Fig.~\ref{Fig1_Str} depicts the unit cell of Lu$_{2}$Ir$_{3}$Si$_{5}$
crystal structure. One observes the absence of transition metal (Ir-Ir)
contacts. The Ir-Si-Ir bond is formed as a cage around Lu atom which
is stacked along the \textit{ab} plane, shown in Fig.~\ref{Fig1_Str}(a).
The Lu atoms form a quasi one - dimensional (1D) zig-zag chain along
the \textit{c}-axis (Fig.~\ref{Fig1_Str}(a)) which are well separated
from the Ir-Si ring. It can be found, in Fig.~\ref{Fig1_Str}(b),
that the zig-zag chain of Lu atom is strongly coupled with the \textit{b}-axis
through Ir1 atom. By analyzing the distances between other atoms,
these Lu atoms have the shortest distance with respect to all other
bonds; suggesting a quasi-1D conducting channel in the Lu - Lu chain,
developing along the \textit{c}-axis. 

\begin{figure}[H]
\centering{} \includegraphics[width=6cm,height=6cm,keepaspectratio]{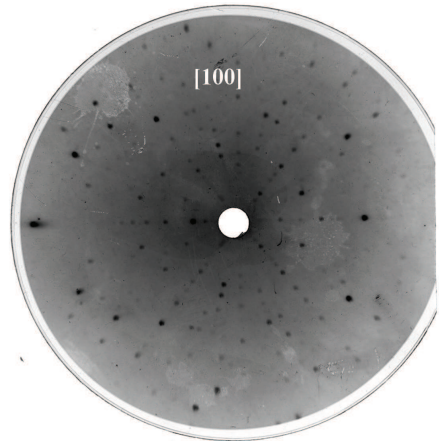}
\caption{\label{Fig2_LP}The observed Laue pattern of Lu$_{2}$Ir$_{3}$Si$_{5}$
along {[}100{]} axis.}
\end{figure}

\begin{figure}[H]
\begin{centering}
\includegraphics[width=8cm,height=8cm,keepaspectratio]{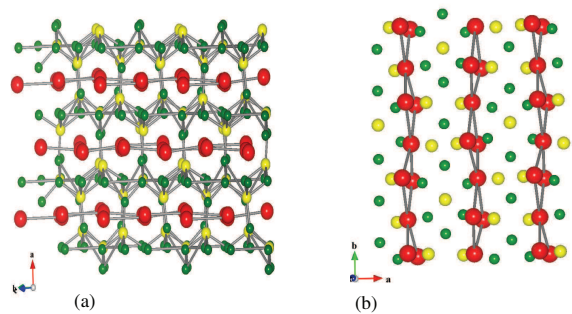}
\par\end{centering}

\caption{\label{Fig1_Str}(Color online) The crystal structure of Lu$_{2}$Ir$_{3}$Si$_{5}$.
The large (red) spheres are correspond to Lu atom, Ir atoms with medium
(yellow) spheres and Si atoms with small (green) spheres.}
\end{figure}

\subsection{Electrical resistivity}

Fig.~\ref{Fig3_R} shows the temperature dependent resistivity of
Lu$_{2}$Ir$_{3}$Si$_{5}$ along\textit{ a, b }and \textit{c} axes.
In this Figure, the inset of the left panel shows $\rho(T)$ taken
during cooling and warming cycles, between 140 and 300~K, at the
rate of 1~K/min. $\rho(T)$ exhibits a sharp upturn between 170 and
250~K, while cooling and warming cycles, signifies the opening up
of a gap in the electronic density of states at the Fermi surface.
This behavior is similar to the one usually observed in charge-density
wave (CDW) transition.During the cooling cycle the upturn of $\rho(T)$,
related to CDW transition, starts at $\thicksim$210~K and reaches
a maximum at $\thicksim$175~K. While in the warming cycle, the maximum
of $\rho(T)$ occurs at $\thicksim$215~K then it decreases sharply
and exhibits minimum at$\thicksim$250~K. 

\begin{figure}[H]
\noindent \centering{}\includegraphics[width=8cm,height=8.5cm]{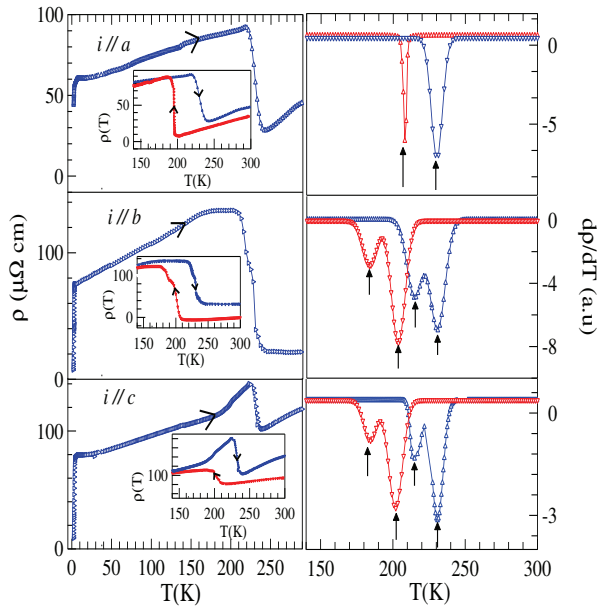}\caption{\label{Fig3_R} (color online) The temperature dependence of the electrical
resistivity $\rho(T)$ of Lu$_{2}$Ir$_{3}$Si$_{5}$. Left panel
shows the resistivity for temperature scans between 1.8 to 300~K.
The inset of left panel shows $\rho(T)$ illustrating the hysteresis
between 140 and 300~K in the resistivity taken on cooling (red triangle)
and warming (blue triangle) cycle for Lu$_{2}$Ir$_{3}$Si$_{5}$
along the \textit{a, b} and \textit{c} axes. Right panel shows \textit{d$\rho$/dT
}as a function of temperature between 140 and 300~K highlights multiple
anomalies. The solid arrows indicate the transition temperatures of
the anomalies.}
\end{figure}

\begin{table}[H]
\begin{centering}
\begin{tabular}{>{\centering}m{1cm}>{\centering}m{3.5cm}>{\centering}m{3.5cm}}
 &  & \tabularnewline
\hline 
\hline 
 &  & \tabularnewline
{\small axis} & {\small Resistivity}{\small \par}

{\small T$_{_{CDW}}$$\mathrm{\left(K\right)}$ } & {\small Susceptibility}{\small \par}

{\small T$_{_{CDW}}$$\mathrm{\left(K\right)}$ }\tabularnewline
\hline 
{\small }%
\begin{tabular}{>{\centering}p{1cm}}
\tabularnewline
{\small a}\tabularnewline
\tabularnewline
{\small b}\tabularnewline
\tabularnewline
{\small c}\tabularnewline
\end{tabular} & {\small }%
\begin{tabular}{>{\centering}p{1.4cm}>{\centering}p{1.5cm}}
{\small Cooling} & \textcolor{black}{\small warming}\tabularnewline
\hline 
 & \tabularnewline
{\small 197} & {\small 231 }\tabularnewline
 & \tabularnewline
{\small 184 }{\small \par}

{\small 200} & {\small 213 }{\small \par}

{\small 232}\tabularnewline
 & \tabularnewline
{\small 184 }{\small \par}

{\small 202} & {\small 215}{\small \par}

{\small 231}\tabularnewline
\end{tabular} & {\small }%
\begin{tabular}{>{\centering}p{1.4cm}>{\centering}p{1.5cm}}
{\small Cooling} & {\small warming}\tabularnewline
\hline 
 & \tabularnewline
{\small 199} & {\small 232}\tabularnewline
 & \tabularnewline
{\small 181}{\small \par}

{\small 201} & {\small 214 }{\small \par}

{\small 231}\tabularnewline
 & \tabularnewline
{\small 183 }{\small \par}

{\small 204} & {\small 212 }{\small \par}

{\small 230}\tabularnewline
\end{tabular}\tabularnewline
 &  & \tabularnewline
\hline 
\hline 
 &  & \tabularnewline
\end{tabular}
\par\end{centering}

\caption{\label{Table_F}CDW transition temperatures $T{}_{CDW}$ observed
from both resistivity and susceptibility measurement techniques. }
\end{table}

As a result of these CDW transitions, a large hysteresis of almost
40-50~K is associated between the up and down scans. This strongly
suggests a first-order CDW transition for the system. Besides, we
can see in the figure that the transition is rather broad one at around
35~K. Such a broad curvature and an unusually large
upward jump seen in the resistivity data with a step like increase
(with decreasing temperature) across the transition signifies the
mixture of multiple CDW phase in the sample (seen in Fig~\ref{Fig3_R})
as were observed in Er$_{5}$Ir$_{4}$Si$_{10}$ \cite{Er5Ir4Si10_SCrystal}.
In order to have a better understanding of CDW transitions, The derivative
of resistivity ($\mathrm{\mathit{d\rho/dT}}$} \textit{vs T}) is also
plotted against temperature (shown in right panel of Fig.~\ref{Fig3_R}).
From this, one can clearly elucidate the presence of multiple transitions
for cooling and warming the sample along \textit{b }and \textit{c
}axes. The anomalies are marked with solid arrows. It must be noted
that there is no second anomaly along the \textit{a-}axis. We shall
return to this point as we discuss further. The characterized CDW
transition temperatures $T{}_{CDW}$ obtained from resistivity measurement
is listed in Table~\ref{Table_F}.

\subsection{Magnetic susceptibility}

Fig.~\ref{Fig4_DC} shows dc magnetic susceptibility and its derivative
of Lu$_{2}$Ir$_{3}$Si$_{5}$, as a function of temperature from
150 to 300~K, with an applied magnetic field \textit{H~}=~5T along
$a,$ $b$ and $c$ axes. 

\begin{figure}[h]
\noindent \centering{}\includegraphics[width=8cm,height=8cm]{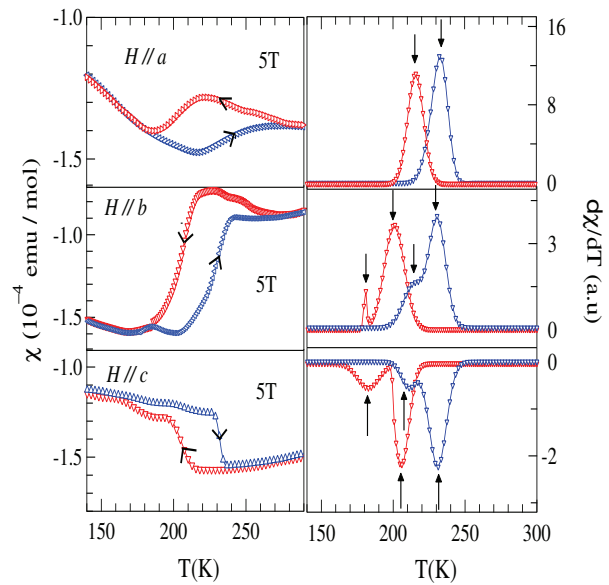}\caption{\label{Fig4_DC} The temperature dependence of the dc susceptibility
$\chi(T)$ of Lu$_{2}$Ir$_{3}$Si$_{5}$. Left panel demonstrates
dc susceptibility in the temperature range between 140 and 300~K
to highlights CDW transition along three principle axes (\textit{a},
\textit{b} and \textit{c}) for both cooling (red triangle) and warming
(blue triangle) the sample, its corresponding derivatives plots are
shown in the right panel of the figure. }
\end{figure}
The magnetic susceptibility data (shown in the left panel) show a
large diamagnetic drop across the phase transition at around 250~K
and 170~K during warming and cooling the sample respectively. This
comes about due to the reduction in the density of states at the Fermi
surface because of the opening up of a gap at the Fermi surface, accompanying
the CDW ordering. A huge thermal hysteresis (40-50~K) associated
with the CDW ordering along the three axes, as observed in our resistivity
results, signifies a first-order characteristic of CDW transition.
The derivative plots of susceptibility ({$\mathrm{\mathit{d\chi/dT}}$
}\textit{vs} \textit{T}), shown in the right panel of Fig.~\ref{Fig4_DC},
demonstrate multiple CDW anomalies for {$b$ and
$c$ axes which are marked by solid arrows. The broad anomalies of
susceptibility data also corroborate the above inference. Concurrently,
one can see a single peak along the $a$ axis as found in resistivity
data.} In contrast, there is an enhancement of magnetization
across the CDW transition along the $c$ axis. We will return to
this point later in the discussion part. The CDW transition temperature
obtained from the susceptibility studies, listed in Table~\ref{Table_F},
are in good agreement with the above mentioned resistivity results.

\subsection{Heat capacity}

Fig.~\ref{Fig5_DSC} shows the zero field specific heat data for
Lu$_{2}$Ir$_{3}$Si$_{5}$ in the temperature range between 150 and
300~K. The background subtraction of specific heat is done by fitting
the lattice contribution of specific heat, for the data far away from
the transition, to demonstrate the heat capacity jumps, $\mathrm{\triangle\mathit{\mathit{C}}_{\mathit{CDW}}}$.
The entropy change $\mathrm{\mathit{\triangle}\mathit{\mathit{S}}_{\mathit{CDW}}}$
across the CDW transition is obtained by integrating the curve under
$\triangle C_{CDW}/T$ as a function of \textit{T.} The transition
temperature \textcolor{black}{$\mathit{T_{CDW}}$}, heat capacity
jumps $\mathrm{\triangle\mathit{\mathit{C}}_{\mathit{CDW}}}$ and
the entropy change $\triangle S$ during warming and cooling the sample
(seen in Fig.~\ref{Fig5_DSC}) are listed in Table~\ref{Table_2}.

\begin{table}[H]
\noindent \begin{centering}
\begin{tabular}{>{\raggedright}m{2.3cm}>{\centering}m{2.2cm}>{\centering}m{1.8cm}>{\centering}m{1.2cm}}
 &  &  & \tabularnewline
\hline 
\hline 
 & {\small $\mathit{T_{_{CDW}}}$ $\mathrm{\left(K\right)}$ } & {\small Total $\triangle C_{_{CDW}}$ (J/ mol K) } & {\small Total}{\small \par}

{\small $\triangle S$(R) }\tabularnewline
\hline 
 &  &  & \tabularnewline
 &  &  & \tabularnewline
{\small }%
\begin{tabular}{>{\centering}p{2.3cm}}
warming curve\tabularnewline
\tabularnewline
cooling curve\tabularnewline
\tabularnewline
\end{tabular} & {\small }%
\begin{tabular}{>{\centering}p{2cm}}
{\small 218~K}{\small \par}

{\small 232~K}\tabularnewline
\tabularnewline
{\small 185~K}{\small \par}

{\small 198~K}\tabularnewline
\tabularnewline
\end{tabular} & {\small }%
\begin{tabular}{>{\centering}p{1cm}}
{\small 62}\tabularnewline
\tabularnewline
{\small 62}\tabularnewline
\tabularnewline
\end{tabular} & {\small }%
\begin{tabular}{>{\centering}p{1cm}}
{\small 0.42}\tabularnewline
\tabularnewline
{\small 0.42}\tabularnewline
\tabularnewline
\end{tabular}\tabularnewline
\hline 
\hline 
 &  &  & \tabularnewline
\end{tabular}
\par\end{centering}

\caption{\label{Table_2} The parameters obtained from heat capacity studies
of Lu$_{2}$Ir$_{3}$Si$_{5}$.}
\end{table}

The values of transition temperatures are in good agreement with both
susceptibility and resistivity results. It has been observed that
the specific heat anomaly and the entropy change for Lu$_{2}$Ir$_{3}$Si$_{5}$
single crystal are much larger and sharper than that of conventional
CDW systems such as K$_{0.3}$MoO$_{3}$ (8~J/mol~K, 0.18\textit{R})
\cite{KmoO3_1,critical_fluctuation_NbSe3,critical_fluctuation_KMoO3}
and NbSe$_{3}$ ($\sim$9~J/mol~K, 0.08\textit{R}) \cite{specific_heat_NbSe3}.

\begin{figure}[h]
\noindent \begin{centering}
\includegraphics[width=8cm,height=8cm,keepaspectratio]{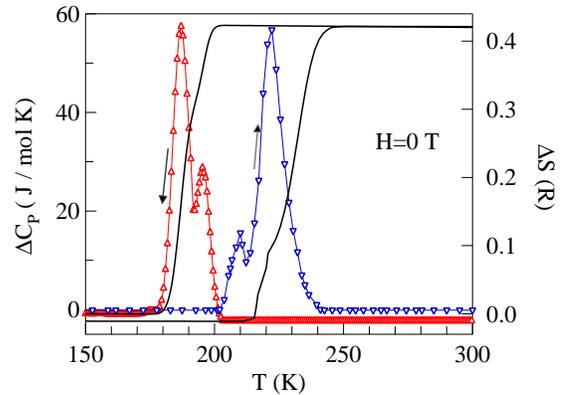}
\par\end{centering}

\noindent \centering{}\caption{\label{Fig5_DSC} The temperature dependence of specific heat measurements
$\triangle C_{P}$ \textit{vs} $T$ (left axis) on both warming (blue
triangle) and cooling (red triangle) of Lu$_{2}$Ir$_{3}$Si$_{5}$
after subtracting the smooth background. Entropy $\triangle S$ associated
with the transition is estimated after background substraction, shown
in right axis of the plot.}
\end{figure}

\begin{figure}[H]
\noindent \centering{}\includegraphics[width=8cm,height=8cm,keepaspectratio]{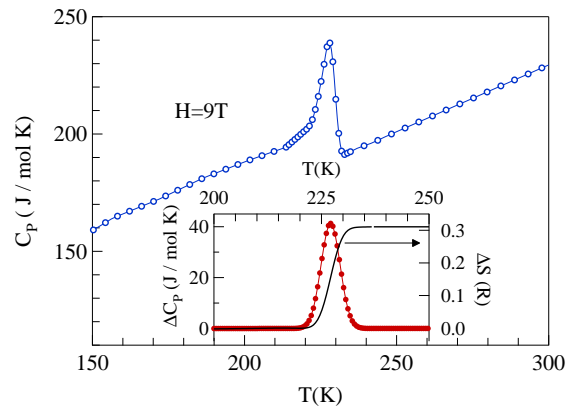}\caption{\label{Figr_PPMS}The specific heat capacities of Lu$_{2}$Ir$_{3}$Si$_{5}$in
the field of 9~T while warming the sample from 150 to 300~K. Inset
shows the specific heat (left axis) on an enlarged temperature scale
and the calculated entropy change (right axis) between 200 and 250~K
.}
\end{figure}

The presence of a sharp peak anomaly in the specific heat data gives
the clear evidence of a high electron density and a large amplitude
of the periodic lattice distortion accompanying the CDW. Compared
to the 2H-TaSe$_{2}$ and 2H-TaS$_{2}$ layered compounds \cite{2H-Tase4},
this large phonon specific heat anomaly may be due to the presence
of an incommensurate CDW phase. The huge cusp in heat capacity data
($\mathrm{\triangle\mathit{\mathit{C}}_{\mathit{CDW}}}$=62 J/mol
K) and the pronounced thermal hysteresis, in the warm up and cooling
down scans, are the characteristic features expected in a strongly
coupled first order CDW transition. The additional peaks observed
at 218~K and at 186~K in the heating and cooling curves, respectively,
(shown in Fig.~\ref{Fig5_DSC}) indicate the mixture of multiple
CDW transition in the compound. These findings are well consistent
with the above mentioned resistivity and susceptibility studies. 

Fig.~\ref{Figr_PPMS} demonstrates the heat capacity data between
150 and 300~K performed in PPMS set up by applying a magnetic field
of 9~T. A huge peak ($\triangle C_{P}$ is 42~J/mol~{K)}
is observed at 232~K, while no anomaly is found at lower transition
218~K. The sharp ($\triangle T_{_{CDW}}/T_{_{CDW}}$$\thicksim$
3\%) upper transition is accompanied by an entropy change of 0.32\textit{R,}
where \textit{R} is a gas constant.

\section{Discussion}

In this section, we will discuss the possible origin
of multiple CDW transitions observed in Lu$_{2}$Ir$_{3}$Si$_{5}$
single crystal. In this context it is worthwhile to recapitulate multiple
CDW transitions observed in other materials.  In the layered compound
NbSe$_{3}$ which shows a two dimensional structure, there are reports
of two CDW transitions occurring at 145 and 49~K \cite{NbSe3_strucuture}.
The transition at 145 K is related to the rearrangement of the atoms
in the four-chain units resulting in the appearance of the CDW phase.
Similarly the transition at 49 K is related to the two-chain units
in the \textit{bc} plane of the crystal. Hence,
the two CDW would have different wavelengths and their formation would
take place at different temperatures and they are not correlated.
So in NbSe$_{3},$ it requires two independent
Fermi nesting conditions for the development of two CDW anomalies.
Multiple CDW transitions is also reported in a few of the R$_{5}$Ir$_{4}$Si$_{10}$
(R=rare earth) compounds \cite{multipleCDW_5-4-10}. In these compounds,
one of the rare earth atoms forms a chain-like structure along the
\textit{c -} axis resulting in the formation of a quasi 1D chain.
The rearrangement of this quasi 1D chain results in the formation of
a CDW phase. It is observed that in these materials the low temperature
purely commensurate CDW phase is achieved via a 1D incommensurate
CDW phase transition at high temperatures. This is endorsed by superlattice
reflections at various temperatures. However this multiple CDW ordering
is not observed in {Lu$_{5}$Ir$_{4}$Si$_{10}$.}

We have demonstrated that Lu$_{2}$Ir$_{3}$Si$_{5}$
exhibits the multiple first order CDW transition by the resistivity,
susceptibility and specific heat measurement. By looking the crystal
structure of Lu$_{2}$Ir$_{3}$Si$_{5},$ (ref. Fig.~\ref{Fig1_Str}),
it can be safely assumed that the CDW anomalies is related to the
quasi 1D chain of Lu atoms along the \textit{c
-} axis. Hence we speculate that this Lu chain
of atoms might be rearranging themselves twice via commensurate/incommensurate
modes, resulting in appearance of multiple CDW phases. Synchrotron
measurements are needed to confirm this. Such multiple transitions
is favourable to be observed along \textit{b}
and \textit{c} axis as this
compound has the zig-zag chain of Lu atoms spread over \textit{bc}
plane. 

Interestingly, the susceptibility data along the
\textit{c}-axis shows an upward
jump across the CDW ordering whereas it shows a drop in susceptibility
along the \textit{a}and \textit{b}
axes. This means that the DOS along the \textit{c}-axis
reduces whereas along \textit{a} and \textit{b} axes it enhances
across the transition. Such an effect was earlier observed in Tm$_{5}$Ir$_{4}$Si$_{10}$
\cite{multipleCDW_5-4-10}. We speculate that Lu atoms play a definite
role in these transitions and further understanding requires band
structure calculations to determine the possible nesting and gaping
of the FS.

\section{conclusion}

To conclude, the detailed bulk measurements suggest the multiple CDW
anomalies in Lu$_{2}$Ir$_{3}$Si$_{5}$ single crystal. In addition,
a large thermal hysteresis of about 50~K has been observed in magnetic
susceptibility, electrical resistivity, and heat capacity measurements
across the CDW ordering which suggests a first order phase transition
associated with incommensurate CDW transition. Besides, the giant
excess of specific heat $\triangle C_{P}/C_{P}\sim26\%$ and the huge
specific heat jump (62~J/mol~K) further support the strong coupling
first order CDW scenario. However, definite conclusion of aforesaid
scenarios need further experimentation on Lu$_{2}$Ir$_{3}$Si$_{5}$,
concerning structural fluctuation and lattice
softening (inelastic neutron scattering). Synchrotron X-ray study of the system is to be performed to determine q vectors for the CDW transition.

\section{ACKNOWLEDGMENT}

N.S.S. thank to Prof. A. K. Grover, TIFR Mumbai, for his various kind
of support throughout the current research work. N.S.S. also grateful
to Bhanu Joshi for his help during the experiments. C. V. Tomy would
like to acknowledge the Department of Science and Technology for the
partial support of this work.


\begin{thebibliography}{References}
\bibitem{Peierls_book}R. E. Peierls, \textit{Quantum Theory of Solids}
(Oxford University Press, New York, 1955).

\bibitem{Gruner}G. Gruner,\textit{ Density Waves in Solids }(Addison-Wesley,
Reading, 1994).

\bibitem{KMoO3}L. Degiorgi, B. Alavi, G. Mihaly, and G. Gruner, Phys.
Rev. B \textbf{44}, 7808 (1991).

\bibitem{Nb_Tb_Se3_1}S. Sridhar, D. Reagor, and G. Gruner, Phys.
Rev. Lett. \textbf{55}, 1196 (1985).

\bibitem{Bechgaard_salt}T. Giamarchi, S. Biermann, A. Georges, and
A. Lichtenstein, J. Phys. IV France \textbf{114}, 23 (2004).

\bibitem{TTF-TCNQ2}A. Schwartz, M. Dressel, G. Gruner, V. Vescoli,
L. Degiorgi, and T. Giamarchi, Phys. Rev. B \textbf{58}, 1261 (1998).

\bibitem{PrPt2Si2_19}{[}34{]}. M. Kumar, V. K. Anand, C. Geibel,
M. Nicklas and Z. Hossain, Phys. Rev. B 81, 125107 (2010).

\bibitem{R2Ir3Si5_Yogesh_18}Yogesh Singh, D. Pal and S. Ramakrishnan,
Phys. Rev. B \textbf{70}, 064403 (2004).

\bibitem{R5Ir4Si10_14}H. D. Yang, P. Klavins, and R. N. Shelton,
Phys. Rev. B 43, 7688 (1991).

\bibitem{R5Ir4Si10_15} K. Ghosh, S. Ramakrishnan, and G. Chandra,
Phys. Rev. B 48, 4152 (1993). 

\bibitem{multipleCDW_5-4-10}S. V. Smaalen, M. Shaz, L. Palatinus,
P. Daniels, F. Galli, G. J. Nieuwenhuys, and J. A. Mydosh,Phys. Rev.
B \textbf{69}, 014103 (2004).

\bibitem{SCrystal_Lu5Ir4Si10}B. Becker, N. G. Patil, S. Ramakrishnan,
A. A. Menovsky, G. J. Nieuwenhuys, J. A. Mydosh, M. Kohgi, and K.
Iwasa, Phys. Rev. B \textbf{59}, 7266 (1999).

\bibitem{lu2Ir3Si5_yogesh}Yogesh Singh, Dilip Pal, S. Ramakrishnan,
A. M. Awasthi, and S. K. Malik, Phys. Rev. B 71, 045109 (2005).

\bibitem{Lu2r3Si5_3}Y. K. Kuo and K. M. Sivakumar, T. H. Su and C.
S. Lue, Phys. Rev. B \textbf{74}, 045115 (2006).

\bibitem{Sangu}N. S. Sangeetha, A. Thamizhavel, C. V Tomy, Saurabh
Basu, A. M. Awasthi, S. Ramakrishnan and D. Pal, Phys. Rev. B \textbf{86},
024524 (2012).

\bibitem{Fullprof}Juan Rodriguez-Carvajal, Phys. B (Amsterdam) \textbf{55},
192 (1993).

\bibitem{Er5Ir4Si10_SCrystal}F. Galli, S. Ramakrishnan, T. Taniguchi,
G. J. Nieuwenhuys, J. A. Mydosh, S. Geupe, J. L\:{u}decke and S.
van Smaalen, Phys. Rev. Lett. \textbf{85}, 158 (2000).

\bibitem{KmoO3_1}J. W. Brill, M. Chung, Y.-K. Kuo, X. Zhan, and E.
Figueroa, Phys. Rev. Lett. 74, 1182 (1995).

\bibitem{critical_fluctuation_KMoO3}R. S. Kwok and S. E. Brown, Phys.
Rev. Lett. \textbf{63}, 895–898 (1989).

\bibitem{critical_fluctuation_NbSe3}R. S. Kwok and G. Gruner, S.
E. Brown, Phys. Rev. Lett. \textbf{65}, 365–368 (1990).

\bibitem{specific_heat_NbSe3}S. Tomić, K. Biljaković, D. Djurek,
J.R. Cooper, P. Monceau, A. Meerschaut, Solid State Commun. 38 109
(1981) .

\bibitem{2H-Tase4} R. A. Craven and S. F. Meyer, Phys. Rev. B \textbf{16},
4583 (1977).

\bibitem{NbSe3_strucuture}J. L Hodeau, M. Marezio, C. Roucau, R.
Ayroles, A. Meerschaut, J. Rouxel and P. Monceau, J. Phys. C: Solid
State Phys., \textbf{11} (1978).\end{thebibliography}
\end{document}